\documentclass[aps,twocolumn,showpacs,superscriptaddress]{revtex4}
\usepackage{graphicx}
\usepackage{amsmath}
\usepackage{amssymb}
\usepackage{color}
\usepackage{braket}
\usepackage{bbm}
\usepackage{bm}
\usepackage{mathrsfs}

\definecolor{gold}{rgb}{1,0.75,0}

\begin{document}

\title{Boson pair production in arbitrarily polarized electric fields}
\author{Z. L. Li \footnote{zlli@cumtb.edu.cn}}
\affiliation{School of Science, China University of Mining and Technology, Beijing 100083, China}
\affiliation{State Key Laboratory for GeoMechanics and Deep Underground Engineering, China University of Mining and Technology, Beijing 100083, China}
\author{B. S. Xie \footnote{Corresponding author. bsxie@bnu.edu.cn}}
\affiliation{Key Laboratory of Beam Technology of the Ministry of Education $\&$ College of Nuclear Science and Technology, Beijing Normal University, Beijing 100875, China}
\affiliation{Beijing Radiation Center, Beijing 100875, China}
\author{Y. J. Li \footnote{Corresponding author. lyj@aphy.iphy.ac.cn}}
\affiliation{School of Science, China University of Mining and Technology, Beijing 100083, China}
\date{\today}

\begin{abstract}
The momentum spectrum and number density of created bosons for two types of arbitrarily polarized electric fields are calculated and compared with those of created fermions, employing the equal-time Feshbach-Villars-Heisenberg-Wigner formalism which is confirmed that for an uniform and time-varying electric field it is completely equivalent to the quantum Vlasov equation in scalar QED. For an elliptically polarized field, it is found that the number density of created bosons is a square root of the number density of spin-up electrons times that of spin-down ones for a circularly polarized multicycle field. Moreover, the degree of spin polarization roughly grows as the Keldysh adiabaticity parameter increases for arbitrarily polarized multicycle fields. For a field constituted of two circularly polarized fields with a time delay, it is shown that momentum vortices also exist in boson pair creation and are induced only by the orbital angular momentum of particles. However, the vortices can reproduce the quantum statistic effect due to the effect of spin of particles. These results further deepen the understanding of some significant signatures in pair production.
\end{abstract}


\maketitle

\section{Introduction}
The phenomenon that quantum vacuum in the presence of a strong field becomes unstable and decays into electron-positron pairs is a fascinating theoretical prediction of quantum electrodynamics (QED) \cite{Sauter,Heisenberg,Schwinger}, which has not been observed experimentally due to the need for high enough field strength, i.e., the external field strength should be comparable to the critical one $E_{\mathrm{cr}}\sim10^{16}\mathrm{V/cm}$, corresponding to laser intensity $I_{\mathrm{cr}}\sim10^{29}\mathrm{W/cm^2}$. Although current laser intensity $\sim10^{22}\mathrm{W/cm^2}$ is far less than the critical field strength, some laser facilities under construction are expected to achieve $\sim10^{26}\mathrm{W/cm^2}$ \cite{ELI,XCELS}. With the rapid development of laser technology, the critical laser intensity is constantly being approached in future, which arouses great interesting to the study of vacuum pair production in strong external fields \cite{Piazza2012,Xie2017}.

Recently many important signatures in pair production for complex but realistic fields are discovered by common methods including wordline instanton technique \cite{Dunne2005,Dunne2006,Dumlu2011}, quantum Vlasov equation (QVE) \cite{Kluger1998,Schmidt1998,Bloch1999}, Wentzel-Kramers-Brillouin (WKB) approximation \cite{Kim2002,Kim2007,CKDumlu2011}, Dirac-Heisenberg-Wigner formalism \cite{Bialynicki1991,Hebenstreit2010,Ababekri2019}, and computation quantum field theory \cite{Su2019,Wang2019}, such as the dynamically assisted Schwinger mechanism \cite{Schutzhold2008,Piazza2009,Orthaber2011,Nuriman2012}, quantum statistics effect \cite{Hebenstreit2009,Dumlu2010}, Ramsey interferences \cite{Akkermans2012,ZLLi2014}, spin polarization in elliptically polarized fields \cite{Wollert2015,Strobel2015,Blinne2016}, effective mass signatures, momentum vortices and ponderomotive effects in multiphoton pair production \cite{Kohlfurst2014,ZLLi2017,Kohlfurst2018,ZLLi2019}, and so on. However, most of them are found in fermion pair production, and it is not sure whether they still exist in boson pair production. In particular, some signatures are changed in the latter case. For instance, the expression of the effective mass at the $n$th threshold given in \cite{Kohlfurst2014} must be modified, because the number density varying with the field frequency for $n$-photon absorption in fermion pair production is similar to that for $(n-1)$-photon absorption in boson pair production. On the another hand, the study of boson pair production can also be used to deepen the understanding of some signatures which can not be fully understood in fermion pair production, such as the formation reason of momentum vortices.

In this paper, we investigate the boson pair production in two types of arbitrarily polarized electric fields using the equal-time Feshbach-Villars-Heisenberg-Wigner (FVHW) formalism, and clarify the following problems: (a) The relation between the FVHW formalism for a spatially homogeneous and time-dependent electric field and the QVE in scalar QED; (b) The effect of field polarizations on boson pair creation and the relation between the number density of created bosons and those of created spin-up and spin-down electrons; (c) The effect of field parameters on spin polarization; (d) The possibility of the existence of momentum vortices in boson pair creation and their formation reason; (e) The effect of spin of particles on the vortices. These studies will further deepen our understanding of relevant signatures in vacuum pair production for arbitrarily polarized electric fields.

This paper is organized as follows: In Sec.~\ref{sec2} we derive the
FVHW formalism for a spatially homogeneous and time-varying electric field and relate it to the QVE in scalar QED. In Sec.~\ref{sec3} we show the numerical results of pair production in a single elliptically polarized field and a field constituted of two circularly polarized fields with a time delay. In Sec.~\ref{sec4} we give our
concludes. To clearly see the relation between the FVHW formalism and the QVE, the QVE in scalar QED is briefly derived in
App.~\ref{appa}. Additionally, the expansion of the Wigner function in scalar QED into eigenoscillations is shown in App.~\ref{appb}.

\section{Theoretical formula}
\label{sec2}

In this section, we will briefly review the equal-time Feshbach-Villars-Heisenberg-Wigner (FVHW) formalism which corresponds to the Wigner function in scalar QED and relate it to the quantum Vlasov equation. Our starting point is the Klein-Gordon equation in an electromagnetic field
\begin{equation}\label{KG}
(D^\mu D_\mu+m^2)\psi(\mathbf{x},t)=0,
\end{equation}
where $D_\mu=\partial_\mu+iqA_\mu(\mathbf{x},t)$ is the covariant derivative, $q$ and $m$ are the particle charge and mass, respectively, and $A_\mu(\mathbf{x},t)$ is the four-dimensional vector potential of the electromagnetic field. Note that here we start from classical fields but quantize the matter field and treat the electromagnetic field as a classical one in the following subsection. Moreover, natural units $\hbar=c=1$ are used throughout this paper.

Using the Feshbach-Villars representation \cite{Feshbach1958}, Equation (\ref{KG}) can be written in Schr\"{o}dinger form
\begin{equation}\label{FV}
i\frac{\partial}{\partial t}\Psi(\mathbf{x},t)=\hat{\mathbbm{H}}(\mathbf{x},t)\Psi(\mathbf{x},t),
\end{equation}
where the two-component Feshbach-Villars field
\begin{eqnarray}\label{FVPsi}
\Psi&=&\left(
                     \begin{array}{c}
                       \xi \\
                       \eta \\
                     \end{array}
                   \right),\\
\xi&=&\frac{1}{2}\Big(\psi+\frac{i}{m}\frac{\partial\psi}{\partial t}-\frac{qA^0}{m}\psi\Big),\\
\eta&=&\frac{1}{2}\Big(\psi-\frac{i}{m}\frac{\partial\psi}{\partial t}+\frac{qA^0}{m}\psi\Big),
\end{eqnarray}
and the two-component nonhermitian Hamiltonian
\begin{equation}\label{Hamiltonian}
\hat{\mathbbm{H}}=\frac{(\hat{p}-q\mathbf{A})^2}{2m}\mathbbm{a}+m\mathbbm{b}+qA^0\mathbbm{1}
\end{equation}
with $\hat{p}=-i\nabla$, $\mathbbm{a}=\sigma_3+i\sigma_2=
\left(
\begin{array}{cc}
    1 & 1 \\
    -1 & -1 \\
\end{array}
\right), \mathbbm{b}=\sigma_3=\left(
  \begin{array}{cc}
    1 & 0 \\
    0 & -1 \\
  \end{array}
\right), \mathbbm{1}=\left(
  \begin{array}{cc}
    1 & 0 \\
    0 & 1 \\
  \end{array}
\right)$. Here $\sigma_1=\left(
  \begin{array}{cc}
    0 & 1 \\
    1 & 0 \\
  \end{array}
\right), \sigma_2=\left(
  \begin{array}{cc}
    0 & -i \\
    i & 0 \\
  \end{array}
\right), \sigma_3=\left(
  \begin{array}{cc}
    1 & 0 \\
    0 & -1 \\
  \end{array}
\right)$ are the Pauli matrices.

\subsection{Derivation of the FVHW formalism}
\label{sec2a}

Before showing the Wigner function, we first give the equal-time density operator which is a equal-time anticommutator constructed by two Feshbach-Villars field operators in the Heisenberg picture:
\begin{eqnarray}
\hat{\varrho}(\mathbf{x}_1,\mathbf{x}_2,t)&\equiv&
    \exp\Big(-iq\int_{\mathbf{x}_2}^{\mathbf{x}_1}
  d\mathbf{x}'\cdot\mathbf{A}(\mathbf{x}',t)\Big) \nonumber \\
  &&\times\{\Psi(\mathbf{x}_1,t),{\Psi}^\dagger(\mathbf{x}_2,t)\} ,
\end{eqnarray}
where the first factor on the right hand side of this equation is a Wilson line factor introduced for preserving gauge invariance. Note that although the integration path in the Wilson line factor is not unique, the choice of the integration path should ensure that the variable $\mathbf{p}$ in the definition of Wigner function Eq. (\ref{wignerfunction}) has a proper physical interpretation. Therefore,  a straight line path which can identify the variable $\mathbf{p}$ as the kinetic momentum is the perfect choice \cite{Elze1986}. In the center of mass coordinate $\mathbf{x}=(\mathbf{x}_1+\mathbf{x}_2)/2$ and the relative coordinate $\mathbf{s}=\mathbf{x}_1-\mathbf{x}_2$, the density operator becomes
\begin{eqnarray}
  \hat{\varrho}(\mathbf{x},\mathbf{s},t)
  &=&\exp\Big(-iq\int_{-1/2}^{1/2}d\lambda\,\mathbf{s}\cdot\mathbf{A}(\mathbf{x}
  +\lambda\mathbf{s},t)\Big) \nonumber\\ &&\times\Big\{\Psi\Big(\mathbf{x}
  +\frac{\mathbf{s}}{2},t\Big),\Psi^\dagger\Big(\mathbf{x}-\frac{\mathbf{s}}{2},t\Big)\Big\}.
\end{eqnarray}

The Wigner function is generally defined as the vacuum expectation value of the Fourier transform of $\hat{\varrho}(\mathbf{x},\mathbf{s},t)$ with respect to the relative coordinate $\mathbf{s}$:
\begin{eqnarray}\label{wignerfunction}
\mathcal{W}(\mathbf{x},\mathbf{p},t)\equiv \frac{1}{2}\int{d^3s\,
\langle\mathrm{vac}|\hat{\varrho}(\mathbf{x},\mathbf{s},t)|\mathrm{vac}\rangle e^{-i\mathbf{p}\cdot\mathbf{s}}},
\end{eqnarray}
where $|\mathrm{vac}\rangle$ is the vacuum state in the Heisenberg picture and $\mathbf{p}$ is the kinetic momentum. Taking the time derivative of Eq.~(\ref{wignerfunction}) and employing Eq. (\ref{FV}), one can derive the equation of motion for the Wigner function as
\begin{eqnarray}\label{EOM}
iD_t\mathcal{W}&=&-i\frac{\mathbf{P}\cdot\mathbf{D}}{2m}(\mathbbm{a}\mathcal{W}
+\mathcal{W}\mathbbm{a}^\dagger) \nonumber\\
&&+
\frac{4\mathbf{P}^2-\mathbf{D}^2}{8m}(\mathbbm{a}\mathcal{W}
-\mathcal{W}\mathbbm{a}^\dagger)\\
&&+m(\mathbbm{b}\mathcal{W}-\mathcal{W}\mathbbm{b}), \nonumber
\end{eqnarray}
where $D_t$, $\mathbf{D}$ and $\mathbf{P}$ denote non-local pseudo-differential operators:
\begin{eqnarray}\label{pseudo-differential}
  D_t&=&\frac{\partial}{\partial t}+q\int_{-1/2}^{1/2}{d\lambda\,\mathbf{E}\Big(\mathbf{x}
  +i\lambda\frac{\partial}{\partial \mathbf{p}},t\Big)\cdot \frac{\partial}{\partial \mathbf{p}}}, \nonumber \\
  \mathbf{D}&=&\nabla+q\int_{-1/2}^{1/2}{d\lambda\,\mathbf{B}\Big(\mathbf{x}
  +i\lambda\frac{\partial}{\partial \mathbf{p}},t\Big)\times\frac{\partial}{\partial \mathbf{p}}}, \\
  \mathbf{P}&=&\mathbf{p}-iq\int_{-1/2}^{1/2}{d\lambda\,\lambda\,\mathbf{B}\Big(\mathbf{x}
  +i\lambda\frac{\partial}{\partial \mathbf{p}},t\Big)\times\frac{\partial}{\partial \mathbf{p}}}. \nonumber
\end{eqnarray}
For more details, see \cite{Best1993PRD,Best1993AoP,Zhuang1998}. Note that a Hartree approximation of the electromagnetic field is adopted in the derivation, namely, the electromagnetic field is regarded as a C-number rather than a Q-number.

To proceed, we may expand the Wigner function in terms of the Feshbach-Villars spinors $\{\mathbbm{1},\sigma_{i=\{1,2,3\}}\}$ and $4$ real functions $\chi^{\mu=\{0,1,2,3\}}(\mathbf{x},\mathbf{p},t)$ which are later called FVHW functions as
\begin{equation}\label{SpinorDeco}
  \mathcal{W}(\mathbf{x},\mathbf{p},t)=\frac{1}{2}\left(\chi^0\mathbbm{1}+\chi^1\sigma_1
  +\chi^2\sigma_2+\chi^3\sigma_3\right).
  \
\end{equation}
Inserting this decomposition into Eq.~(\ref{EOM}) and comparing the coefficients of the Feshbach-Villars spinors, we obtain a set of partial differential equations for the $4$ FVHW functions:
\begin{eqnarray}\label{Decos}
D_t\chi^0\!&=&\!\frac{4\mathbf{P}^2-\mathbf{D}^2}{4m}\chi^2
-\frac{\mathbf{P}\cdot\mathbf{D}}{m}\chi^3,\\
D_t\chi^1\!&=&\!-\Big(\frac{4\mathbf{P}^2-\mathbf{D}^2}{4m}+2m\Big)\chi^2
+\frac{\mathbf{P}\cdot\mathbf{D}}{m}\chi^3,\\
D_t\chi^2\!&=&\!\frac{4\mathbf{P}^2-\mathbf{D}^2}{4m}\chi^0+
\Big(\frac{4\mathbf{P}^2-\mathbf{D}^2}{4m}+2m\Big)\chi^1, \;\\
D_t\chi^3\!&=&\!-\frac{\mathbf{P}\cdot\mathbf{D}}{m}(\chi^0+\chi^1).
\end{eqnarray}
The above equations, along with Eq. (\ref{pseudo-differential}), are called the FVHW formalism. Some of the FVHW functions can be given an obvious physical interpretation by the expression of conservation laws for some physically observable quantities, such as the total charge $\mathcal{Q}$, the total current $\mathcal{J}$, the total energy $\mathcal{E}$, and the total linear momentum $\mathcal{P}$, i.e.,
\begin{equation}
  \frac{d}{dt}\left\{\mathcal{Q};\mathcal{J};\mathcal{E};\mathcal{P}\right\}=0 \ ,
\end{equation}
with
\begin{eqnarray}
\mathcal{Q} \!&=&\! q\int d\Gamma\, \chi^3(\mathbf{x},\mathbf{p},t), \\
\mathcal{J} \!&=&\! q\int d\Gamma\, \frac{\mathbf{p}}{m}[\chi^0(\mathbf{x},\mathbf{p},t)+\chi^1(\mathbf{x},\mathbf{p},t)], \\
\mathcal{E} \!&=&\! \int d\Gamma\, \Big[\Big(\frac{\mathbf{p}^2}{2m}+m\Big)\chi^0(\mathbf{x},\mathbf{p},t)
  +\frac{\mathbf{p}^2}{2m}\chi^1(\mathbf{x},\mathbf{p},t)\Big] \nonumber\\
  &&+\frac{1}{2}\int d^3x\, [\mathbf{E}^2(\mathbf{x},t)+\mathbf{B}^2(\mathbf{x},t)], \\
\mathcal{P} \!&=&\! \int d\Gamma\, \mathbf{p}\,\chi^3(\mathbf{x},\mathbf{p},t)+\int d^3x\, \mathbf{E}(\mathbf{x},t)\times\mathbf{B}(\mathbf{x},t),
\end{eqnarray}
where $d\Gamma=d^3x\,d^3p/(2\pi)^3$. According to the above expressions, $m \chi^0(\mathbf{x},\mathbf{p},t)$ may be associated with a mass density, $\frac{\mathbf{p}}{m}[\chi^0(\mathbf{x},\mathbf{p},t)+\chi^1(\mathbf{x},\mathbf{p},t)]$ with a current density and $q \chi^3(\mathbf{x},\mathbf{p},t)$ with a charge density. However, the quantity $\chi^2(\mathbf{x},\mathbf{p},t)$ has no obvious physical interpretation.

\subsection{Relating the FVHW formalism to QVE}
\label{sec2b}

In this subsection we will show that the FVHW formalism for a spatially homogeneous and time-varying electric field $\mathbf{E}(\mathbf{x},t)=\mathbf{E}(t)$ is equivalent to the well-known QVE in scalar QED.

First, we determine the initial conditions of the FVHW formalism by calculating the Wigner function for vanishing external fields $A_\mu(\mathbf{x},t)=0$. The Feshbach-Villars field operator can be expressed in terms of the bosonic annihilation operator of particles $\hat{a}_\mathbf{p}$ and creation operator of antiparticles $\hat{b}^\dagger_{-\mathbf{p}}$ as
\begin{equation}\label{FVSolution}
\Psi(\mathbf{x},t)=\!\!\int\!\frac{d^3p}{(2\pi)^3}\big[u_+(\mathbf{p},t)\hat{a}_\mathbf{p}
+u_-(-\mathbf{p},t)\hat{b}^\dagger_{-\mathbf{p}}\big]e^{i\mathbf{p}\cdot\mathbf{x}}, \!\!
\end{equation}
where $u_\pm(\mathbf{p},t)=u_\pm(\mathbf{p})e^{\mp i\omega(\mathbf{p})t}$,
\begin{equation}\label{FVSolution1}
u_\pm(\mathbf{p})=\frac{1}{2}\left(
                    \begin{array}{c}
                      \sqrt{m/\omega(\mathbf{p})}\pm \sqrt{\omega(\mathbf{p})/m}\\
                      \sqrt{m/\omega(\mathbf{p})}\mp \sqrt{\omega(\mathbf{p})/m}\\
                    \end{array}
                  \right)
\end{equation}
denotes the particles with positive and with negative energy, respectively, and $\omega(\mathbf{p})=\sqrt{\mathbf{p}^2+m^2}$ is the particle energy. Substituting Eqs. (\ref{FVSolution}) and (\ref{FVSolution1}) into Eq. (\ref{wignerfunction}), we derive the Wigner function for pure vacuum:
\begin{equation}
\mathcal{W}_{\mathrm{vac}}(\mathbf{x},\mathbf{p},t)=\frac{1}{4}\left(
     \begin{array}{cc}
       \frac{m}{\omega(\mathbf{p})}+\frac{\omega(\mathbf{p})}{m} & \frac{m}{\omega(\mathbf{p})}-\frac{\omega(\mathbf{p})}{m} \\
       \frac{m}{\omega(\mathbf{p})}-\frac{\omega(\mathbf{p})}{m} & \frac{m}{\omega(\mathbf{p})}+\frac{\omega(\mathbf{p})}{m} \\
     \end{array}
   \right).
\end{equation}
By comparing the above expression with Eq.~(\ref{SpinorDeco}), we finally obtain the vacuum initial conditions:
\begin{eqnarray}\label{initialvalues}
\chi_{\mathrm{vac}}^0(\mathbf{p})&=&\frac{1}{2}\Big[\frac{m}{\omega(\mathbf{p})}
+\frac{\omega(\mathbf{p})}{m}\Big]\ , \nonumber \\
\chi_{\mathrm{vac}}^1(\mathbf{p})&=&\frac{1}{2}\Big[\frac{m}{\omega(\mathbf{p})}
-\frac{\omega(\mathbf{p})}{m}\Big]\ , \\
\chi_{\mathrm{vac}}^2(\mathbf{p})&=&\chi_{\mathrm{vac}}^3(\mathbf{p})=0.\nonumber
\end{eqnarray}

Next let us give the FVHW formalism for a spatially homogeneous and time-dependent electromagnetic field with vanishing magnetic field. In this case, the non-local operators Eq.~(\ref{pseudo-differential}) become local ones:
\begin{eqnarray}
  D_t&=& \frac{\partial}{\partial  t} + q \mathbf{E}(t)\cdot\frac{\partial}{\partial \mathbf{p}} , \nonumber \\
  \mathbf{D}&=& \nabla, \\
  \mathbf{P}&=&\mathbf{p}. \nonumber
\end{eqnarray}
Moreover, since the FVHW functions are not the function of variable $\mathbf{x}$, all spatial derivatives vanish. As a consequence, according to the vacuum initial conditions Eq. (\ref{initialvalues}), we find that the function $\chi^3(\mathbf{p},t)\equiv0$ and the FVHW formalism is reduced to
\begin{equation}\label{PDEs}
\left[\frac{\partial}{\partial t}+q\mathbf{E}(t)\cdot\frac{\partial}{\partial \mathbf{p}}\right]\left\{
                    \begin{array}{c}
                      \chi^0 \\
                      \chi^1 \\
                      \chi^2 \\
                    \end{array}
                  \right\}(\mathbf{p},t)
=\mathcal{M}(\mathbf{p})\left\{
                    \begin{array}{c}
                      \chi^0 \\
                      \chi^1 \\
                      \chi^2 \\
                    \end{array}
                  \right\}(\mathbf{p},t),
\end{equation}
where
\begin{equation}
 \mathcal{M}(\mathbf{p})=\left(\begin{array}{ccc}
 0&0&\frac{\mathbf{p}^2}{m}\\
 0&0&-\frac{\mathbf{p}^2}{m}-2m\\
 \frac{\mathbf{p}^2}{m}&\frac{\mathbf{p}^2}{m}+2m&0
\end{array}\right) \
\end{equation}
is a $3\times3$ matrix.

Applying the method of characteristics \cite{Hebenstreit2010}, or simply replacing $\mathbf{p}(t)$ by $\mathbf{k}-q\mathbf{A}(t)$ , the partial differential equations Eq.~(\ref{PDEs}) can be simplified as an ordinary differential equation system:
\begin{eqnarray}\label{ODEs}
\frac{d}{d t} \chi^0(\mathbf{k},t)\!&=&\!\frac{\mathbf{p}^2(t)}{m}\chi^2(\mathbf{k},t),\nonumber\\
\frac{d}{d t} \chi^1(\mathbf{k},t)\!&=&\!-\Big[\frac{\mathbf{p}^2(t)}{m}+2m\Big]\chi^2(\mathbf{k},t),\\
\frac{d}{d t} \chi^2(\mathbf{k},t)\!&=&\!\frac{\mathbf{p}^2(t)}{m}\chi^0(\mathbf{k},t)+
\Big[\frac{\mathbf{p}^2(t)}{m}+2m\Big]\chi^1(\mathbf{k},t). \nonumber \,\,\,\,\quad
\end{eqnarray}
Note that we have chosen the temporal gauge $A_0=0$ here as used in the derivation of QVE, see App. \ref{appa}.

Introducing three auxiliary quantities
\begin{eqnarray}\label{auxi}
\widetilde{\chi}^1\!&=&\!\Big[\frac{\mathbf{p}^2(t)}{2m\omega(\mathbf{k},t)}
+\frac{m}{\omega(\mathbf{k},t)}\Big]\chi^0
+\frac{\mathbf{p}^2(t)}{2m\omega(\mathbf{k},t)}\chi^1, \nonumber\quad\\
\mathcal{G}\!&=&\!\frac{\mathbf{p}^2(t)}{2m\omega(\mathbf{k},t)}\chi^0
+\Big[\frac{\mathbf{p}^2(t)}{2m\omega(\mathbf{k},t)}
+\frac{m}{\omega(\mathbf{k},t)}\Big]\chi^1, \quad\\
\mathcal{H}\!&=&\!\chi^2,\nonumber
\end{eqnarray}
the ODE system Eq.~(\ref{ODEs}) can be transformed into
\begin{eqnarray}\label{FVHW}
  \frac{d}{dt}\mathcal{F}(\mathbf{k},t)\!\!&=&\!\!\frac{1}{2}W(\mathbf{k},t)\,\mathcal{G}(\mathbf{k},t) \ , \nonumber \\
  \!\frac{d}{dt}\mathcal{G}(\mathbf{k},t)\!\!&=&\!\!W(\mathbf{k},t)[1+2\mathcal{F}(\mathbf{k},t)]
  -2\omega(\mathbf{k},t)\,\mathcal{H}(\mathbf{k},t), \;\;\quad
  \\
  \frac{d}{dt}\mathcal{H}(\mathbf{k},t)\!\!&=&\!\!2\omega(\mathbf{k},t)\,\mathcal{G}(\mathbf{k},t) \ , \nonumber
\end{eqnarray}
where $2\mathcal{F}(\mathbf{k},t)=\widetilde{\chi}^1(\mathbf{k},t)-1$ parameterizes the deviation from the vacuum state and vanishes in pure vacuum, and
\begin{equation}
W(\mathbf{k},t)=\frac{q\mathbf{E}(t)\cdot\mathbf{p}(t)}{\omega^2(\mathbf{k},t)}.
\end{equation}
It is worth noting that this transformation is roughly equivalent to expanding Wigner function into eigenoscillations \cite{Best1993PRD,Best1993AoP}, see also App. \ref{appb}. Additionally, according to Eqs. (\ref{initialvalues}) and (\ref{auxi}), we derive the initial conditions for Eq. (\ref{FVHW}): $\mathcal{F}(\mathbf{k},t\rightarrow-\infty)=\mathcal{G} (\mathbf{k},t\rightarrow-\infty)=\mathcal{H}(\mathbf{k},t\rightarrow-\infty)=0$. Refer to Eq. (\ref{QVE2}) in App. \ref{appa}, one can clearly see that the Eq. (\ref{FVHW}) is nothing but the well-known QVE in its differential form. Finally, we show that the FVHW formalism for a spatially homogeneous and time-dependent electric field is fully equivalent to QVE. Of course, the FVHW formalism is a more general approach than QVE for studying vacuum pair production in more complex and realistic fields.

The number density of created real boson pairs can be calculated by integrating the momentum distribution function $\mathcal{F}(\mathbf{k},t\rightarrow\infty)$ with respect to the momentum $\mathbf{k}$:
\begin{equation}\label{numberdensity}
 n(\mathbf{k},t\rightarrow\infty)=\int \frac{d^3k}{(2\pi)^3}\mathcal{F}(\mathbf{k},t\rightarrow\infty).
\end{equation}

\section{Numerical results}\label{sec3}

From Eq. (\ref{FVHW}), it is easy to see that this equation (or the QVE in scalar QED) can be used to study boson pair production from vacuum in a spatially homogeneous and time-dependent electric field with three components. However, the QVE in spinor QED applies only to the study of fermion pair creation in one-component electric field. To study the fermion pair production in the electric field with three components, one can use the Dirac-Hensenberg-Wigner formalism given in \cite{Blinne2014}. In this section, we investigate the momentum spectra and number density of created boson pairs in two types of arbitrarily polarized electric fields, and compare the results with those in the case of fermion pair production.

\subsection{Single elliptically polarized field}\label{sec3a}

The first field we considered is a single elliptically polarized electric field:
\begin{eqnarray}\label{Field1}
\mathbf{E}(t)=E_1\,e^{-\frac{t^2}{2\tau^2}}\left(
                                             \begin{array}{c}
                                               \cos(\omega_1 t) \\
                                               \delta_1\sin(\omega_1 t) \\
                                               0 \\
                                             \end{array}
                                           \right),
\end{eqnarray}
where $E_{1}=E_{01}/\sqrt{1+\delta_1^2}$ is the field amplitude, $\tau$ is the pulse duration, $\omega_{1}$ is the field frequency, and $\delta_{1}\in[-1,1]$ is the field ellipticity. Considering the symmetry of the dependency of the following results on the ellipticity, we only choose $0\leq\delta_1\leq1$ in this subsection.

\begin{figure*}[!ht]
  \centering
  \includegraphics[width=\textwidth]{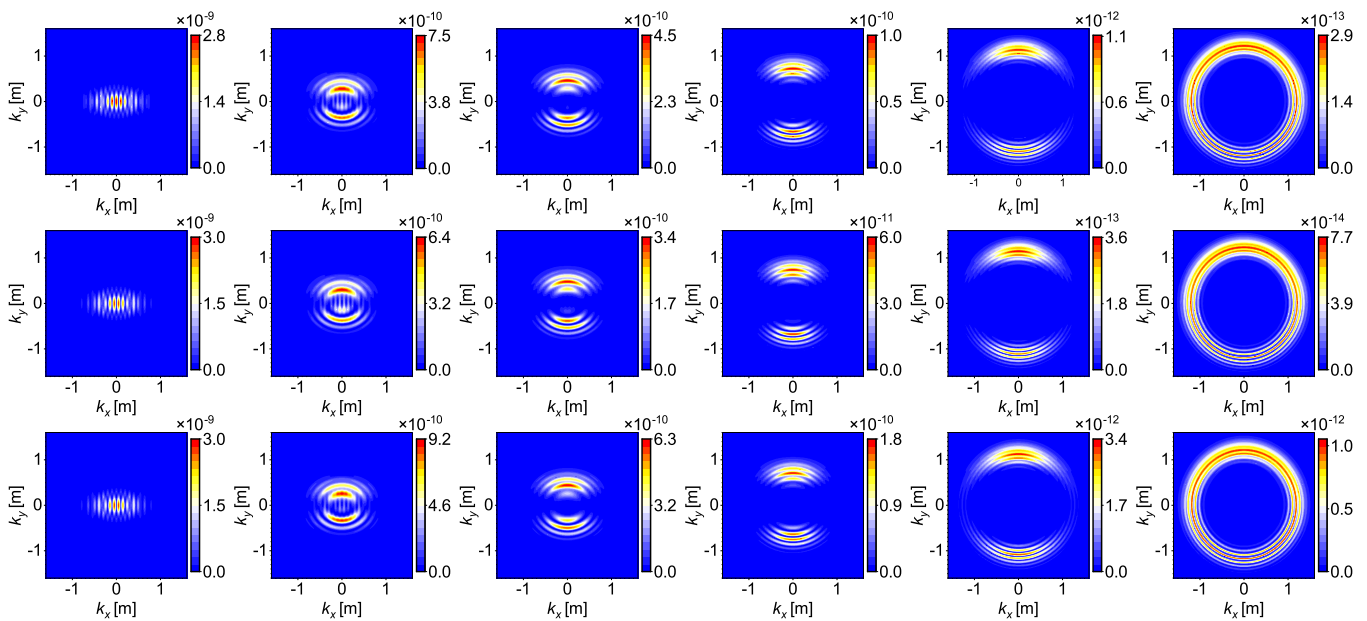}\\
  \caption{Momentum spectra of created particles in the polarized plane ($k_z=0$) for an elliptically polarized electric field with different ellipticities. For each row, from left to right, $\delta_1=0,\,0.2,\,0.3,\,0.5,\,0.9$ and $1$. The three rows, from top to bottom, correspond to bosons, spin-up electrons and spin-down electrons, respectively. Other field parameters are $E_{01}=0.1\sqrt{2}\,E_\mathrm{cr}$, $\omega=0.1m$ and $\tau=100/m$.}\label{Fig1}
\end{figure*}

In the first row of Fig. \ref{Fig1}, we show the momentum spectra of created bosons in the polarized plane ($k_x$, $k_y$, $k_z=0$) for an elliptically polarized electric field with different polarizations. To compare the results of boson and fermion pair production, we also depict the momentum spectra of spin-up electrons (second row) and spin-down electrons (third row) in Fig. \ref{Fig1}. Note that the electron charge and mass are used in the comparison, the results of fermion pair creation are calculated by the method used in \cite{Blinne2016}, and the spin quantization direction is along the $z$ axis (more descriptions see \cite{Description}). For a linearly polarized field $\delta_1=0$, one can see that the momentum spectrum has obvious interference pattern along the direction of electric field, and the maximum and minimum positions of the distribution function is complete opposite to those in fermion pair production which is an embodiment of the quantum statistics effect \cite{Hebenstreit2009,ZLLi2014}. For a nonlinearly polarized field $\delta_1\neq0$, with the increase of ellipticity, the momentum spectrum is first split into two parts along the direction of momentum $k_y$, and then connected into a ring. This result is similar to that in fermion pair creation \cite{ZLLi2015}, see also the last two rows in Fig. \ref{Fig1}. Another signature of the momentum spectrum for $\delta_1\neq0$ is that the spectrum is symmetry along the direction of the kinetic momentum $k_x$ while it is asymmetry along the direction of $k_y$. This phenomenon is associated with the parity of electric field components, because the electric field component $E_x(t)$ is an even function of time which leads to the total energy of created particles satisfying $\omega(-k_x,k_y,k_z,-t)=\omega(k_x,k_y,k_z,t)$ and finally produces a symmetric spectrum, however, $E_y(t)$ is an odd function of time which generates an asymmetric spectrum. For more detailed discussions, see the first paragraph above Sec. V on page 8 in \cite{CKDumlu2011}. Furthermore, the maximum and minimum positions of interference fringes are roughly the same as those for fermions, which is different from the result for a linearly polarized field. Mathematically, this is because the turning point structures (complex form solutions to $\omega(\mathbf{k},t)=0$ for a fixed momentum $\mathbf{k}$) in the complex $t$ plane for linearly polarized fields are changed by the nonlinearly polarized fields \cite{Olugh2019}. For example, the interchange of the oscillatory minima and maxima of the momentum spectrum between bosons ($+$) and fermions ($-$) can be clearly seen, if there are two pair of turning points closest to the real time axis, from the expression $\mathcal{F}(\mathbf{k})\approx e^{-2K_1}+e^{-2K_2}\pm2\cos(2\alpha)e^{-K_1-K_2}$ derived by the phase integral method, where $K_1=\big|\int_{t_1}^{t_1^*}\omega(\mathbf{k},t)dt\big|,\, K_2=\big|\int_{t_2}^{t_2^*}\omega(\mathbf{k},t)dt\big|$, and $t_1,t_1^*,t_2,t_2^*$ are two different pairs of complex conjugate turning points \cite{Dumlu2010}. However, the interchange will vanish if only one pair of turning points dominates as $\mathcal{F}(\mathbf{k})\approx e^{-2K_1}$. That is, the change of the turning point structures can result in the absence of the quantum statistics effect. Physically, since the photons in an elliptically polarized field carry spin angular momentum, the particles created by absorbing multiple photons can obtain orbital angular momentum based on the conservation of angular momentum \cite{ZLLi2019}, which weaken the effect of spin on the positions of interference fringes \cite{ZLLi20152}. Of course, for the pair production dominated by Schwinger tunneling mechanism, the created particles does not obtain the orbital angular momentum from external fields, so that the quantum statistics effect is not affected.

\begin{figure}[!ht]
  \centering
  \includegraphics[width=8.5cm]{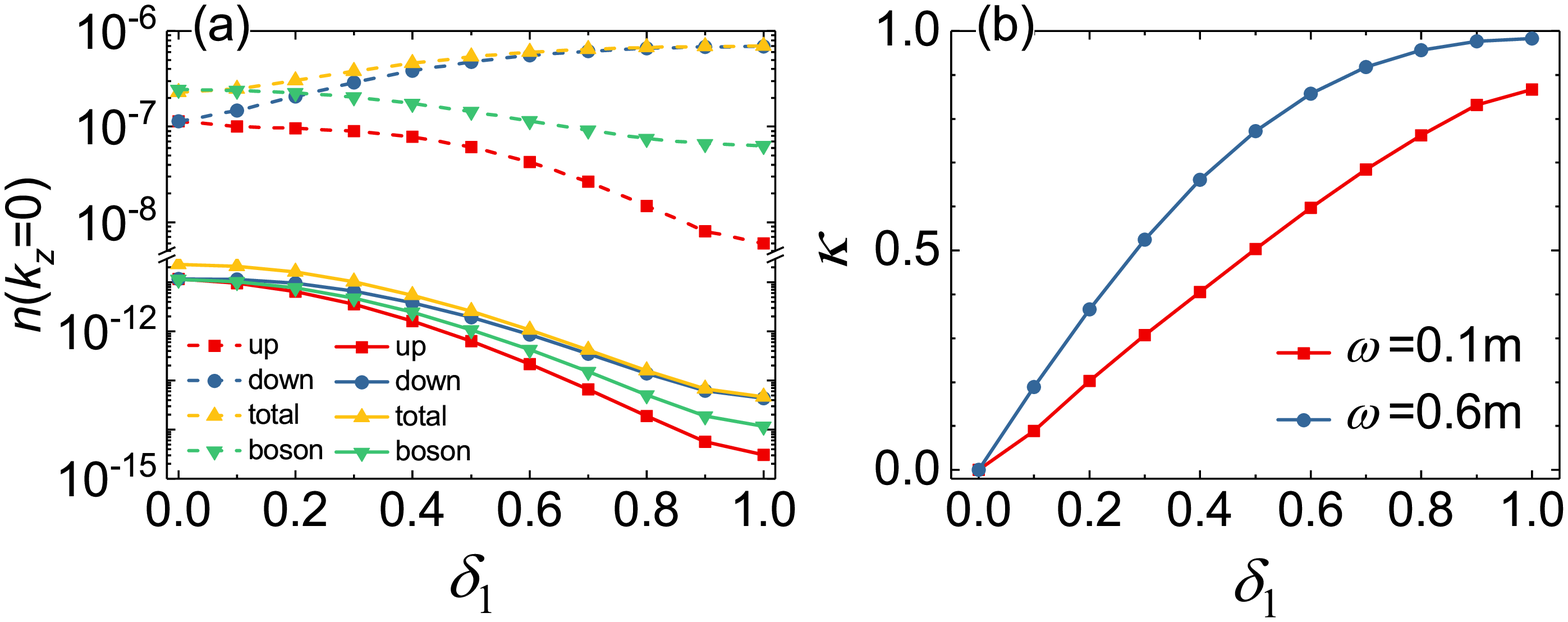}\\
  \caption{(a) is the number density of created particles in the polarized plane ($k_z=0$) as a function of the ellipticity for $\omega=0.1m$ (solid line) and $\omega=0.6m$ (dashed line), and (b) shows the degree of spin polarization $\kappa$ varying with the ellipticity. Other field parameters are $E_{01}=0.1\sqrt{2}\,E_\mathrm{cr}$ and $\tau=100/m$. }\label{Fig2}
\end{figure}

\begin{figure*}[!ht]
  \centering
  \includegraphics[width=\textwidth]{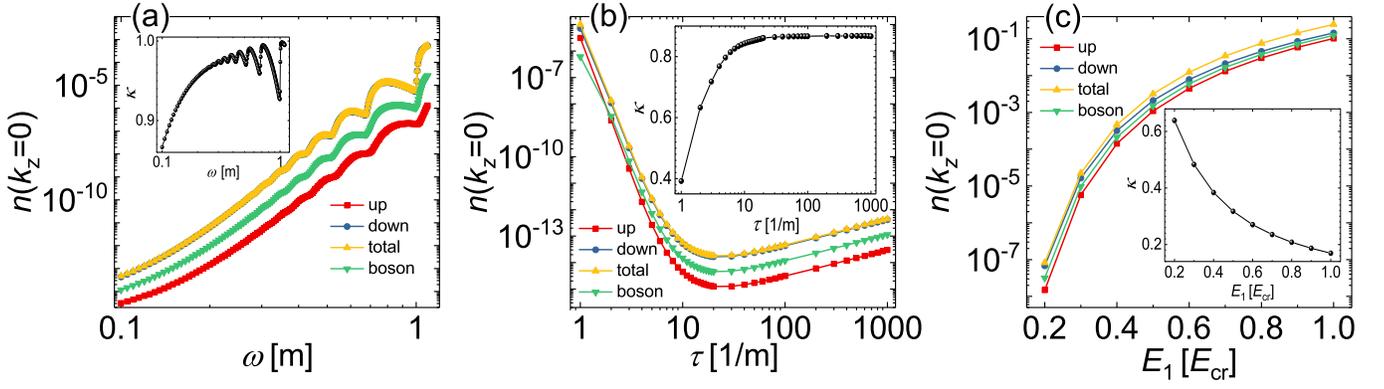}\\
  \caption{The number density of created particles in the polarization plane for a circularly polarized field ($\delta_1=1$) as a function of the field frequency (a) , the pulse duration (b) and the field amplitude (c) . The insets show the DOSP $\kappa$ varying with field parameters. Other field parameters are (a) $E_{01}=0.1\sqrt{2}\,E_\mathrm{cr}$ and $\tau=100/m$, (b) $E_{01}=0.1\sqrt{2}\,E_\mathrm{cr}$ and $\omega=0.1m$, and (d) $\omega=0.1m$ and $\tau=100/m$. }\label{Fig3}
\end{figure*}

By analyzing the value of distribution function, it is also found that its maximum value decreases with the increase of the ellipticity for all cases. This gives us a signal that the particle yield may decrease with the ellipticity as well. To see this clearly, we show the number density of created particles in the polarization plane changing with the ellipticity for the field frequency $\omega=0.1m$ and $\omega=0.6m$ in Fig. \ref{Fig2} (a). One can see that the number density for both bosons and fermions indeed decrease with the ellipticity for $\omega=0.1m$. However, although the number density of created bosons and spin-up electrons still decrease with the ellipticity for $\omega=0.6m$, the number density of created spin-down electrons and total electrons increase with the ellipticity. More calculations show that the optimal number density of created bosons and fermions only corresponds to a linearly polarized field for a small field frequency, while could correspond to, a linearly polarized, a circularly polarized even an elliptically polarized field for a large field frequency. The latter result is a perfect embodiment of the effective mass model failure, because the threshold frequencies for $n$-photon pair production still change with the ellipticity nonlinearly, though the effective mass $m_*=m[1+(mE_{01}/E_{\mathrm{cr}})^2/2]^{1/2}$ is the same for arbitrarily polarized electric fields, see Fig. 5 in \cite{ZLLi2015}. Therefore, for a given field frequency the number density may decrease or increase or even increase first and then decrease with the ellipticity increasing. More interestingly, we find that the number density of created bosons $n_{\mathrm{boson}}$ is approximately equal to the square root of the number density of created spin-up electrons $n_{\mathrm{up}}$ times that of created spin-down electrons $n_{\mathrm{down}}$ for any ellipticity in the case of a small field frequency (such as $\omega=0.1m$):
\begin{equation}\label{b-u-d}
 n_{\mathrm{boson}}\approx\sqrt{n_\mathrm{up}\cdot n_\mathrm{down}}.
\end{equation}
In the case of a large field frequency, for instance $\omega=0.6m$, however, the above relation is broken for small ellipticities while still holds true for a circularly polarized field $\delta_1=1$. More generally, the relation Eq. (\ref{b-u-d}) holds true for any circularly polarized field with sufficient oscillation periods, see Fig. \ref{Fig3}.

Additionally, it is found that the difference between the number density of created spin-up electrons and spin-down electrons becomes larger as the ellipticity increases, and finally the latter one dominates the pair creation process. This is a spin polarization effect  \cite{Blinne2016,Strobel2015,Sorbo2018}, and may be a new way to produce longitudinally polarized relativistic electron (positron) beams which are widely used in high-energy physics. For the convenience of elaboration, we define the degree of spin polarization (DOSP) as
\begin{equation}
\kappa=\frac{n_\mathrm{down}-n_\mathrm{up}}{n_\mathrm{down}+n_\mathrm{up}}.
\end{equation}
In \cite{Blinne2016,Strobel2015}, it shows that the DOSP decreases with the increase of the pulse duration $\tau$ for a fixed parameter $\sigma=\omega\tau$ which is associated with the number of cycles of external fields. However, in this case, the field frequency $\omega$ also varies with the change of the pulse duration $\tau$. So it cannot determine how field parameters affect on the DOSP. To find out, we first consider the effect of the ellipticity on the DOSP, see Fig. \ref{Fig2} (b), where shows the DOSP varying with the ellipticity for $\omega=0.1m$ and $0.6m$. It can be seen that the DOSP $\kappa$ grows with the ellipticity for both cases, and it is always greater for $\omega=0.6m$ than for $\omega=0.1m$. Particularly, for $\omega=0.6m$, the DOSP approaches $1$ for a circularly polarized field. In Fig. \ref{Fig3}, we further study the effect of the field frequency (a), the pulse duration (b), and the field amplitude (c) on the DOSP, and find that the DOSP increases with the field frequency except for some oscillations, decreases with the field amplitude, and is little affected by a large pulse duration. All of these results can be summarized as follows: The DOSP roughly grows with the increase of the Keldysh adiabaticity parameter $\gamma=m\omega/|qE_1|$ \cite{Keldysh1964} for an elliptically polarized electric field with sufficient oscillation periods. Here it should be noted that since the field amplitude $E_1=E_{01}/\sqrt{1+\delta_1^2}$ decreases as the ellipticity increases, the above conclusion includes the effect of the ellipticity on the DOSP as well.

%

\subsection{Two circularly polarized field with a time delay}\label{sec3b}

The second field we considered is a electric field constituted of two circularly polarized field with a time delay:
\begin{eqnarray}\label{Field2}
\mathbf{E}(t)&=&E_1\,e^{-\frac{t^2}{2\tau^2}}\left(
                                             \begin{array}{c}
                                               \cos(\omega_1 t) \\
                                               \delta_1\sin(\omega_1 t) \\
                                               0 \\
                                             \end{array}
                                           \right) \nonumber \\
&&+E_2\,e^{-\frac{(t-T)^2}{2\tau^2}}\left(
                                             \begin{array}{c}
                                               \cos\big(\omega_2 (t-T)\big) \\
                                               \delta_2\sin\big(\omega_2 (t-T)\big) \\
                                               0 \\
                                             \end{array}
                                           \right),\quad\;\;
\end{eqnarray}
where $E_{2}=E_{02}/\sqrt{1+\delta_2^2}$ and $\omega_{2}$ are the field amplitude and frequency of the second field, respectively, $\delta_{2}=\pm\delta_1=\pm1$ is the field ellipticity, and $T$ is the time delay between these two rotating fields.

\begin{figure}[!ht]
  \centering
  \includegraphics[width=8.5cm]{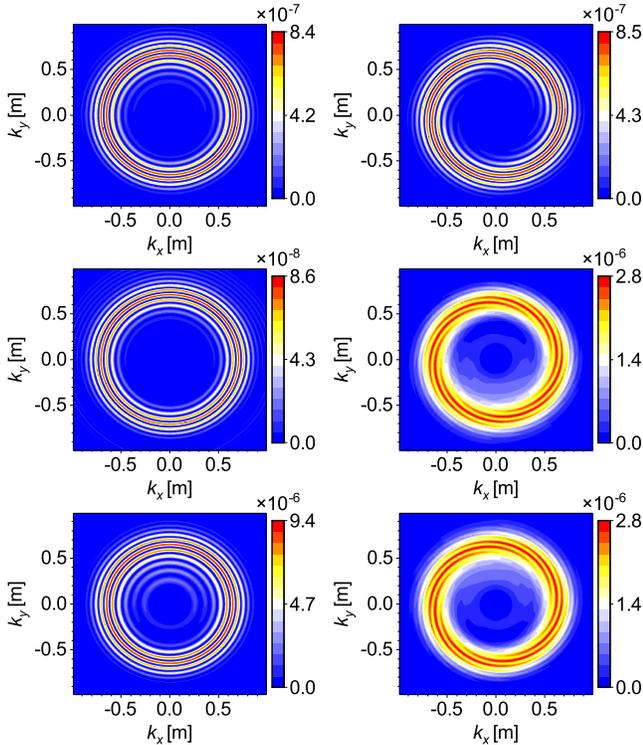}\\
  \caption{Momentum spectra of created particles in the polarized plane ($k_z=0$) for two co-rotating circularly polarized fields ($\delta_1=\delta_2=1$, first column) and counter-rotating ones ($\delta_1=-\delta_2=1$, second column) with a time delay $T=100/m$. The first row corresponds to bosons. The second and third rows correspond to spin-up and spin-down electrons, respectively. Other field parameters are $E_{01}=E_{02}=0.1\sqrt{2}\,E_{\mathrm{cr}}$, $\omega_1=\omega_2=0.6m$ and $\tau=10/m$.}\label{Fig4}
\end{figure}

\begin{figure*}[!ht]
  \centering
  \includegraphics[width=16cm]{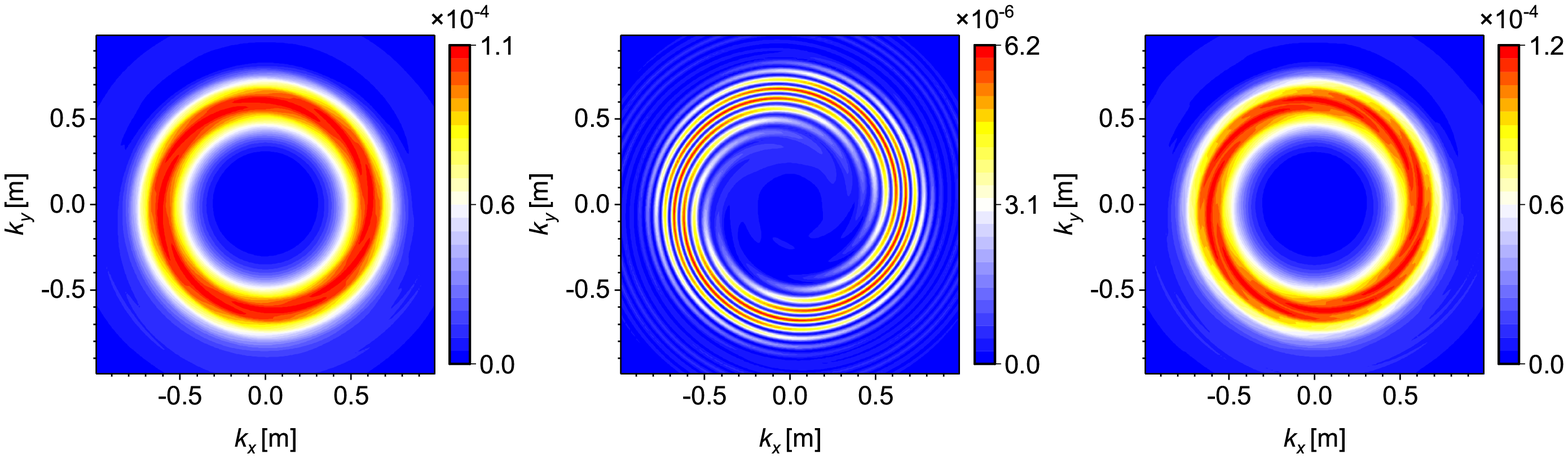}\\
  \caption{Momentum spectra of created spin-up, spin-down and total electrons in the polarized plane ($k_z=0$), from left to right, for two counter-rotating circularly polarized fields ($\delta_1=-\delta_2=1$). The field parameters are $E_{01}=0.1\sqrt{2}E_{\mathrm{cr}},\,E_{02}=0.17\sqrt{2}E_{\mathrm{cr}}$, $\omega_1=\omega_2=0.6m$, $\tau=10/m$ and $T=100/m$.}\label{Fig5}
\end{figure*}

The momentum spectra of created bosons in the polarized plane  ($k_x$, $k_y$, $k_z=0$) for two co-rotating fields ($\delta_1=\delta_2=1$, first column) and counter-rotating ones ($\delta_1=-\delta_2=1$, second column) are shown in the first row of Fig. \ref{Fig4}. For comparison, the momentum spectra of created spin-up and spin-down electrons are also shown in the second and third rows, respectively. One can see that the momentum spectra of created bosons present obvious concentric rings for two co-rotating fields and an eight-start spiral vortex pattern for two counter-rotating fields, which are similar to those in the case of fermion pair production \cite{ZLLi2019,ZLLi2017}. The first result is a common Ramsey interference, but the quantum statistics effect is absent because the interference fringes can also be well determined by Eq. (6) in \cite{ZLLi2017}. That is to say, the spin of particles has little effect on the positions of interference fringes in the case of co-rotating fields. The second result is characterized as a nontrivial Ramsey interference, and the vortex pattern is formed by the interference of the created particle wave packets which have different orbital angular momenta obtained from the spin angular momentum carried by the two counter-rotating fields via multiphoton absorption. Due to the fact that the boson we considered here is a spin-$0$ particle, we finally confirm that the vortex structure is determined by the orbital angular momentum of created particles rather than the spin angular momentum. This result can also be supported by comparing the momentum spectra of created spin-up (second row and second column in Fig. \ref{Fig4}) and spin-down electrons (third row and second column in Fig. \ref{Fig4}). Surprisingly, in the case of counter-rotating fields, the quantum statistics effect is reproduced, i.e., the maxima of interference fringes for bosons correspond to the minima for fermions, and vise versa. Intuitively, it seems that the spin of particles can rotate the vortex pattern around its center by an angle $(2i+1)\pi/\ell$, where $i=0, \pm1,\pm2, ...$,  and $\ell$ is the total number of photons absorbed in multiphoton pair creation or the number of spirals in a vortex. For instance, the minima of rotation angle are $\pm\pi/8$ for the eight-start spiral vortex shown in Fig. \ref{Fig4}. From the point of view of the semiclassical analysis, the spin of particles contribute an odd multiple of $\pi$ to the accumulated phase between the two counter-rotating fields \cite{ZLLi2014,Strobel2015,ZLLi2017}. Moreover, because of the spin polarization effect, the values of momentum distribution functions are also changed by the spin of particles for both cases.

Another issue that needs to be discussed is the minimum values of interference fringes for the vortex structures of created fermions. From the vortex structures of created spin-up and spin-down electrons shown in Fig. \ref{Fig4}, one can see that the minima of interference fringes are nonzero, which is different from that in the case of boson pair production. To analyze the issue clearly, we depict the momentum spectra of created spin-up, spin-down and total electrons in Fig. \ref{Fig5}. Note that the amplitudes of two counter-rotating fields are different from each other in order to ensure that the created spin-down electrons for the second rotating field are comparable to those for the first one. From Fig. \ref{Fig5}, we clearly see that there is no obvious vortex pattern in the left panel, while there is a vortex whose minima of interference fringes are zero presented in the middle one. This result is caused by the spin polarization effect for an elliptically polarized field and can be explained as follows. In subsection \ref{sec3a}, we know that the number density of created spin-down electrons is much greater than that of created spin-up electrons for a circularly polarized field with $\delta_1=+1$, and the result would be reversed for the field with $\delta_1=-1$. Therefore, the electron wave packet created in the first field can be completely canceled out by the one created in the second field for certain momenta in the case of spin-down electrons, but it cannot in the case of spin-up electrons. Moreover, since the number density of created spin-up electrons dominates the pair production, the vortex structure of created total electrons shown in the right panel is not obvious either.

By the way, although we only study the momentum vortices of created bosons for two circularly polarized electric fields with a time delay, the effects of the ellipticity, the relative carrier envelope phase and the time delay on vortex structures and the odd-start spiral vortices can also be considered. However, as the results for bosons are similar to those for fermions, we do not show them here any more.

\section{Conclusions}\label{sec4}

In this paper, we first derive the equal-time FVHW formalism for a spatially homogeneous and time-dependent electric field, which can be used to study boson pair creation in an uniform and time-varying electric field with three components, and show that it is completely equivalent to the QVE in scalar QED. Then employing this method, the momentum spectrum and number density of created bosons in two types of arbitrarily polarized electric fields are investigated.

For an single elliptically polarized field, it is found that the momentum spectrum and number density of created bosons changing with the ellipticity are similar to those in fermion pair production. By comparing with fermion pair production, we find that the number density of created bosons is a square root of the number density of created spin-up electrons times that of spin-down electrons for arbitrarily polarized multicycle fields with a small field frequency and circularly polarized multicycle fields with a large field frequency. Furthermore, the degree of spin polarization roughly grows with the increase of the Keldysh adiabaticity parameter while has little change with the varying of the pulse duration for any elliptically polarized field with sufficient oscillation periods.

For the field constituted of two circularly polarized electric field with a time delay, we confirm that the vortex pattern also exists in boson pair creation, and is caused by the orbital angular momentum of created particles rather than the spin angular momentum. Comparing with the results in fermion pair creation, it is found that the positions of Ramsey interference fringes formed in two co-rotating fields are not affected by the spin of particles. However, the vortex structure formed in two counter-rotating fields embodies the quantum statistics effect discovered in linearly polarized electric fields, i.e., the maxima of interference fringes of the vortex for bosons correspond to the minima for fermions, and vice versa. In addition, we also find that the reason why the minima of interference fringes of the vortex for fermions are nonzero is because the electron wave packets with a specific spin created in the two counter-rotating fields can not completely cancel out due to the spin polarization effect.

The above results further deepen our understanding of some significant signatures in vacuum pair production, and may provide a theoretical reference for exploring pair creation from vacuum experimentally.

\acknowledgments
The work of Z. L. L. and Y. J. Li. is supported by the National Natural Science Foundation of China (NSFC) under Grant  No. 11705278 and No. 11974419, in part by the National Key R\&D Program of China under Grant No. 2018YFA0404802, and by the Fundamental Research Funds for the Central Universities. The work of B. S. X. is supported by NSFC under Grants No. 11875007 and No. 11935008.

\appendix

\section{Quantum Vlasov Equation}\label{appa}

In this appendix, we briefly derive the quantum Vlasov equation in scalar QED. Using the temporal gauge $A_0=0$, the Klein-Gordon equation (\ref{KG}) reads:
\begin{equation}\label{KG1}
\Big(\frac{\partial^2}{\partial t^2}-[\nabla-iq\mathbf{A}(\mathbf{x},t)]^2+m^2\Big)\psi(\mathbf{x},t)=0.
\end{equation}

For an uniform and time-varying electric field, the vector potential $\mathbf{A}(\mathbf{x},t)$ becomes $\mathbf{A}(t)$ and the momentum $\mathbf{k}$ is a good quantum number. Thus, the field operator $\psi(\mathbf{x},t)$ can be decomposed as
\begin{equation}\label{FourierDeco1}
\psi(\mathbf{x},t)=\int \frac{d^3k}{(2\pi)^3}\big[\varphi(\mathbf{k},t)\hat{a}_\mathbf{k}
+\varphi^*(\mathbf{k},t)\hat{b}^\dagger_{-\mathbf{k}}\big]e^{i\mathbf{k}\cdot\mathbf{x}},
\end{equation}
in terms of the time-independent annihilation operator of particles $\hat{a}_\mathbf{k}$ and the creation operator of antiparticles $\hat{b}^\dagger_{-\mathbf{k}}$ which satisfy standard bosonic commutation relations and $\hat{a}_\mathbf{k}|\mathrm{vac}\rangle=\langle\mathrm{vac}|\hat{b}^\dagger_{-\mathbf{k}}=0$. Here the complex mode functions $\varphi(\mathbf{k},t)$ satisfy
\begin{equation}\label{ModeFunction1}
\Big[\frac{d^2}{d t^2}+\omega^2(\mathbf{k},t)\Big]\varphi(\mathbf{k},t)=0,
\end{equation}
with $\omega^2(\mathbf{k},t)=[\mathbf{k}-q\mathbf{A}(t)]^2+m^2$.

In the presence of an electric field, the Hamiltonian operator corresponding to the decomposition (\ref{FourierDeco1}) has nonzero off-diagonal elements, which indicates the pair creation and pair annihilation. To diagonalize the Hamiltonian operator, one can expand the field operator in the adiabatic number basis as
\begin{equation}\label{FourierDeco2}
\psi(\mathbf{x},t)=\int \frac{d^3k}{(2\pi)^3}\big[\widetilde{\varphi}(\mathbf{k},t)\hat{a}_\mathbf{k}(t)
+\widetilde{\varphi}^*(\mathbf{k},t)\hat{b}^\dagger_{-\mathbf{k}}(t)\big]e^{i\mathbf{k}\cdot\mathbf{x}},
\end{equation}
by the adiabatic mode functions
\begin{equation}\label{ModeFunction2}
\widetilde{\varphi}(\mathbf{k},t)=\frac{1}{\sqrt{2\omega(\mathbf{k},t)}}e^{-i\int^t_{-\infty} \omega(\mathbf{k},t') dt'}
\end{equation}
and the time-dependent annihilation operator for particles $\hat{a}_\mathbf{k}(t)$ and the creation operator for antiparticles $\hat{b}_{-\mathbf{k}}(t)$ which satisfy equal-time commutation relations.

The above two decompositions are associated with each other by a time-dependent Bogoliubov transformation. For the mode functions, the form is
\begin{eqnarray}\label{Relation1}
\varphi(\mathbf{k},t)&\!=\!&\alpha_\mathbf{k}(t)
\widetilde{\varphi}(\mathbf{k},t)+\beta_\mathbf{k}(t)\widetilde{\varphi}^*(\mathbf{k},t), \nonumber \\
\\
\frac{d}{d t}\varphi(\mathbf{k},t)&\!=\!&-i\omega(\mathbf{k},t)\big[\alpha_\mathbf{k}(t)
\widetilde{\varphi}(\mathbf{k},t)-\beta_\mathbf{k}(t)\widetilde{\varphi}^*(\mathbf{k},t)\big],
\qquad \nonumber
\end{eqnarray}
where $\alpha_\mathbf{k}(t)$ and $\beta_\mathbf{k}(t)$ are the time-dependent Bogoliubov coefficients and satisfy $|\alpha_\mathbf{k}(t)|^2-|\beta_\mathbf{k}(t)|^2=1$ which is the consequence of bosonic commutation relations. For the creation/annihilation operators, equivalently, the form is
\begin{eqnarray}\label{Relation2}
\left(
  \begin{array}{c}
    \hat{a}_\mathbf{k}(t) \\
    \hat{b}_{-\mathbf{k}}(t) \\
  \end{array}
\right)=\left(
          \begin{array}{cc}
            \alpha_\mathbf{k}(t) & \beta^*_\mathbf{k}(t) \\
            \beta_\mathbf{k}(t) & \alpha^*_\mathbf{k}(t) \\
          \end{array}
        \right)\left(
                  \begin{array}{c}
                    \hat{a}_\mathbf{k} \\
                    \hat{b}_{-\mathbf{k}} \\
                  \end{array}
                \right).
\end{eqnarray}

From Eqs. (\ref{ModeFunction1}), (\ref{ModeFunction2}) and (\ref{Relation1}), we have
\begin{equation}\label{BogoliubovParameters}
\begin{split}
  \frac{d}{d t}\alpha_\mathbf{k}(t) &= \frac{1}{2}W(\mathbf{k},t)\beta_\mathbf{k}(t)e^{2i\int^t_{-\infty} \omega(\mathbf{k},t') dt'}, \\
  \frac{d}{d t}\beta_\mathbf{k}(t) &= \frac{1}{2}W(\mathbf{k},t)\alpha_\mathbf{k}(t)e^{-2i\int^t_{-\infty} \omega(\mathbf{k},t') dt'}.
\end{split}
\end{equation}

In addition, based on the definition of single-particle distribution function
\begin{equation}
f(\mathbf{k},t)\equiv\langle\mathrm{vac}|\hat{a}^\dagger_\mathbf{k}(t)\hat{a}_\mathbf{k}(t)
|\mathrm{vac}\rangle=|\beta_\mathbf{k}(t)|^2
\end{equation}
and using Eqs. (\ref{BogoliubovParameters}), one can derive the integro-differential form of the quantum Vlasov equation:
\begin{eqnarray}\label{QVE1}
\frac{d}{d t}f(\mathbf{k},t)=\frac{1}{2}W(\mathbf{k},t)\int_{-\infty}^t d t'\, \hspace{-4mm}&&W(\mathbf{k},t')[1+2f(\mathbf{k},t')] \nonumber\\
\hspace{-4mm}&&\times\cos[2\Theta(\mathbf{k},t',t)]
\end{eqnarray}
with $\Theta(\mathbf{k},t',t)=\int_{t'}^{t} d \tau\,\omega(\mathbf{k},\tau)$. Introducing two auxiliary quantities
\begin{equation}
g(\mathbf{k},t)=\!\int_{-\infty}^t d t'\,W(\mathbf{k},t')[1+2f(\mathbf{k},t')]\cos[2\Theta(\mathbf{k},t',t)]\nonumber
\end{equation}
and
\begin{equation}
h(\mathbf{k},t)=\!\int_{-\infty}^t d t'\,W(\mathbf{k},t')[1+2f(\mathbf{k},t')]\sin[2\Theta(\mathbf{k},t',t)],\nonumber
\end{equation}
equation (\ref{QVE1}) can be equivalently transformed into a set of ODEs:
\begin{eqnarray}\label{QVE2}
\frac{d}{d t}f(\mathbf{k},t)&\!=\!&\frac{1}{2}W(\mathbf{k},t)g(\mathbf{k},t), \nonumber\\
\frac{d}{d t}g(\mathbf{k},t)&\!=\!&W(\mathbf{k},t)[1+2f(\mathbf{k},t)]-2\omega(\mathbf{k},t)h(\mathbf{k},t)  , \qquad\;\\
\frac{d}{d t}h(\mathbf{k},t)&\!=\!&2\omega(\mathbf{k},t)g(\mathbf{k},t),\nonumber
\end{eqnarray}
with the initial conditions $f(\mathbf{k},t\rightarrow-\infty)=g(\mathbf{k},t\rightarrow-\infty)=h(\mathbf{k},t\rightarrow-\infty)=0$. Note that the distribution function $f(\mathbf{k},t)$ is only meaningful at $t\rightarrow+\infty$.

\section{Expanding Wigner function into eigenoscillations}\label{appb}

Expanding Wigner function into eigenoscillations:
\begin{equation}\label{1}
\mathcal{W}(\mathbf{p},t)=\frac{1}{2}\sum_{i=1}^4\widetilde{\chi}^i(\mathbf{p},t)\mathbbm{e}_i(\mathbf{p}),
\end{equation}
where $\widetilde{\chi}^{i=\{1,2,3,4\}}(\mathbf{p},t)$ are four complex functions and
\begin{eqnarray}\label{2}
\hspace{-1cm}\mathbbm{e}_1(\mathbf{p})\!&=&\!\frac{1}{2}[u_+(\mathbf{p})\otimes u_+(\mathbf{p})+u_-(\mathbf{p})\otimes u_-(\mathbf{p})]\nonumber\\
            \!&=&\!\frac{1}{2}\left(
                      \begin{array}{cc}
                        \frac{m}{\omega(\mathbf{p})}+\frac{\omega(\mathbf{p})}{m} & \frac{m}{\omega(\mathbf{p})}-\frac{\omega(\mathbf{p})}{m} \\
                        \frac{m}{\omega(\mathbf{p})}-\frac{\omega(\mathbf{p})}{m} & \frac{m}{\omega(\mathbf{p})}+\frac{\omega(\mathbf{p})}{m} \\
                      \end{array}
                    \right),
\end{eqnarray}
\begin{eqnarray}
\hspace{-1cm}\mathbbm{e}_2(\mathbf{p})\!&=&\!\frac{1}{2}[u_+(\mathbf{p})\otimes u_+(\mathbf{p})-u_-(\mathbf{p})\otimes u_-(\mathbf{p})]\nonumber\\
            \!&=&\!\left(
                      \begin{array}{cc}
                        1 & 0 \\
                        0 & -1 \\
                      \end{array}
                    \right),
\end{eqnarray}
\begin{eqnarray}
\mathbbm{e}_{3/4}(\mathbf{p})\!&=&\!u_\mp(\mathbf{p})\otimes u_\pm(\mathbf{p}) \nonumber\\
                \!&=&\!\frac{1}{4}\left(
                      \begin{array}{cc}
                        \frac{m}{\omega(\mathbf{p})}-\frac{\omega(\mathbf{p})}{m} & \frac{m}{\omega(\mathbf{p})}+\frac{\omega(\mathbf{p})}{m}\mp2 \\
                        \frac{m}{\omega(\mathbf{p})}-\frac{\omega(\mathbf{p})}{m}\pm2 & \frac{m}{\omega(\mathbf{p})}-\frac{\omega(\mathbf{p})}{m} \\
                      \end{array}
                    \right)\!\!. \qquad
\end{eqnarray}
where $\otimes$ denotes the outer product. For the spatially homogeneous and time-dependent electric field, the nonzero components $\widetilde{\chi}^{i=\{1,3,4\}}(\mathbf{k},t)$ satisfy
\begin{eqnarray}\label{Decos1}
\frac{d}{dt}\widetilde{\chi}^1(\mathbf{k},t)\!&=&\!W(\mathbf{k},t)\big[\widetilde{\chi}^3(\mathbf{k},t)
+\widetilde{\chi}^4(\mathbf{k},t)\big],\\
\frac{d}{dt}\widetilde{\chi}^3(\mathbf{k},t)\!&=&\!W(\mathbf{k},t)\widetilde{\chi}^1(\mathbf{k},t)
+2i\omega(\mathbf{k},t)\widetilde{\chi}^3(\mathbf{k},t), \;\;\quad\\
\frac{d}{dt}\widetilde{\chi}^4(\mathbf{k},t)\!&=&\!W(\mathbf{k},t)\widetilde{\chi}^1(\mathbf{k},t)
-2i\omega(\mathbf{k},t)\widetilde{\chi}^4(\mathbf{k},t), \;\;\quad
\end{eqnarray}
Note that the method of characteristics is used here.

Suppose $2\widetilde{\mathcal{G}}(\mathbf{k},t)=
\widetilde{\chi}^3(\mathbf{k},t)+\widetilde{\chi}^4(\mathbf{k},t)
$ and $2i\widetilde{\mathcal{H}}(\mathbf{k},t)=
\widetilde{\chi}^3(\mathbf{k},t)-\widetilde{\chi}^4(\mathbf{k},t)$, then we have
\begin{eqnarray}\label{Decos2}
\frac{d}{dt}\widetilde{\chi}^1(\mathbf{k},t)\!&=&\!W(\mathbf{k},t)\widetilde{\mathcal{G}}(\mathbf{k},t),\\
\frac{d}{dt}\widetilde{\mathcal{G}}(\mathbf{k},t)\!&=&\!W(\mathbf{k},t)\widetilde{\chi}^1(\mathbf{k},t)
-2\omega(\mathbf{k},t)\widetilde{\mathcal{H}}(\mathbf{k},t), \\
\frac{d}{dt}\widetilde{\mathcal{H}}(\mathbf{k},t)\!&=&\!
2\omega(\mathbf{k},t)\widetilde{\mathcal{G}}(\mathbf{k},t).
\end{eqnarray}

\end{document}